\documentclass[sigconf]{acmart}

\usepackage{booktabs} % For formal tables
\usepackage[
  disable % uncomment to disable todo, listoftodos
]{todonotes}
\usepackage{xcolor,listings}
\usepackage{lipsum}

\usepackage{amssymb}
\usepackage{tikz}
\usetikzlibrary{calc} % for let
\usetikzlibrary{shapes,snakes} % for let
\usetikzlibrary{decorations.pathreplacing}
\usetikzlibrary{patterns}

\usepackage{enumitem} % to adjust indentation

\usepackage{comment}

\usepackage{multirow}

\usepackage[flushleft]{threeparttable}

\usepackage[htt]{hyphenat}

\usepackage{textcomp} % for \texttrademark

\usepackage{balance} % ensure that columns on last page are same length

\copyrightyear{2024}
\acmYear{2024}
\setcopyright{rightsretained}
\acmConference[SIGMOD-Companion '24]{Companion of the 2024
  International Conference on Management of Data}{June 9--15,
  2024}{Santiago, AA, Chile}
\acmBooktitle{Companion of the 2024 International Conference on
  Management of Data (SIGMOD-Companion '24), June 9--15, 2024,
  Santiago, AA, Chile}
\acmDOI{10.1145/3626246.3653374}
\acmISBN{979-8-4007-0422-2/24/06}

\DeclareSymbolFont{textcomp}{TS1}{\ttdefault}{m}{n}
\DeclareMathSymbol{`}{\mathord}{textcomp}{39}
\DeclareMathSymbol{\mathdblquotechar}{\mathalpha}{letters}{`"}
\newcommand{\mathdblquote}{\mathtt{\mathdblquotechar}}
\begingroup\lccode`~=`"\lowercase{\endgroup
  \let~\mathdblquote
}
\AtBeginDocument{\mathcode`"="8000 }

\begin{document}
\title[Measures in SQL]{Measures in SQL}
\newcommand{\version}{2024/04/11}

% Uncomment the following line to show the version number in the title
% \subtitle{(draft, \version{})}

\author{Julian Hyde}
\orcid{0009-0003-7857-7467}
\affiliation{%
  \institution{Google Inc.}
  \streetaddress{345 Spear Street}
  \city{San Francisco}
  \state{CA}
  \country{USA}
  \postcode{94105}
}
\email{julianhyde@google.com}

\author{John Fremlin}
\orcid{0009-0002-6591-6370}
\affiliation{%
  \institution{Google Inc.}
  \streetaddress{111 Eighth Avenue}
  \city{New York}
  \state{NY}
  \country{USA}
  \postcode{10011}
}
\email{fremlin@google.com}

\begin{abstract}

SQL has attained widespread adoption, but Business Intelligence tools
still use their own higher level languages based upon a
multidimensional paradigm. Composable calculations are what is missing
from SQL, and we propose a new kind of column, called a measure, that
attaches a calculation to a table. Like regular tables, tables with
measures are composable and closed when used in queries.

SQL-with-measures has the power, conciseness and reusability of
multidimensional languages but retains SQL semantics. Measure
invocations can be expanded in place to simple, clear SQL.

To define the evaluation semantics for measures, we introduce
context-sensitive expressions (a way to evaluate multidimensional
expressions that is consistent with existing SQL semantics), a concept
called evaluation context, and several operations for setting and
modifying the evaluation context.

% Strategy for implementing measures and contextual expressions based
% on formula advertisement

% Reusable

\end{abstract}




\maketitle

% Format for SQL. Use \, for spaces, ` for single-quotes, e.g.
%   $\sql{SELECT\,`beer`\,FROM\,t}$.
\newcommand{\sql}[1]{\ensuremath{\mathtt{#1}}}

% Format for simple SQL that should flow.
\newcommand{\ssql}[1]{\texttt{#1}}

% SQL syntax
\newcommand{\aggregate}{\ssql{AGGREGATE}}
\newcommand{\asMeasure}{\ssql{AS} \ssql{MEASURE}}
\newcommand{\at}{\ssql{AT}}
\newcommand{\all}{\ssql{ALL}}
\newcommand{\countSql}{\ssql{COUNT}}
\newcommand{\countStar}{\ssql{COUNT(*)}}
\newcommand{\cube}{\ssql{CUBE}}
\newcommand{\current}{\ssql{CURRENT}}
\newcommand{\depend}{\ssql{DEPEND}}
\newcommand{\eval}{\ssql{EVAL}}
\newcommand{\filter}{\ssql{FILTER}}
\newcommand{\from}{\ssql{FROM}}
\newcommand{\groupBy}{\ssql{GROUP} \ssql{BY}}
\newcommand{\groupingId}{\ssql{GROUPING\textunderscore{}ID}}
\newcommand{\having}{\ssql{HAVING}}
\newcommand{\isNotDistinctFrom}{\ssql{IS} \ssql{NOT} \ssql{DISTINCT} \ssql{FROM}}
\newcommand{\join}{\ssql{JOIN}}
\newcommand{\lastValue}{\ssql{LAST\textunderscore{}VALUE}}
\newcommand{\matchRecognize}{\ssql{MATCH\textunderscore{}RECOGNIZE}}
\newcommand{\measure}{\ssql{MEASURE}}
\newcommand{\on}{\ssql{ON}}
\newcommand{\overSql}{\ssql{OVER}}
\newcommand{\partitionBy}{\ssql{PARTITION\,BY}}
\newcommand{\select}{\sql{SELECT}}
\newcommand{\set}{\sql{SET}}
\newcommand{\visible}{\sql{VISIBLE}}
\newcommand{\where}{\sql{WHERE}}
\newcommand{\withinDistinct}{\ssql{WITHIN} \ssql{DISTINCT}}
\newcommand{\withinGroup}{\ssql{WITHIN} \ssql{GROUP}}

% SQL object names
\newcommand{\computeSumRevenue}{\ssql{computeSumRevenue}}
\newcommand{\cost}{\ssql{cost}}
\newcommand{\orderDate}{\ssql{orderDate}}
\newcommand{\enhancedCustomers}{\ssql{Enhanced\-Customers}}
\newcommand{\enhancedOrders}{\ssql{EnhancedOrders}}
\newcommand{\orders}{\ssql{Orders}}
\newcommand{\prodName}{\ssql{prodName}}
\newcommand{\custName}{\ssql{custName}}
\newcommand{\profitMargin}{\ssql{profitMargin}}
\newcommand{\customers}{\ssql{Customers}}
\newcommand{\custAge}{\ssql{custAge}}
\newcommand{\revenue}{\ssql{revenue}}
\newcommand{\sumRevenue}{\ssql{sumRevenue}}
\newcommand{\summarizedOrders}{\ssql{SummarizedOrders}}
\newcommand{\orderYearSql}{\ssql{orderYear}}

\lstset{language=SQL,
       showspaces=false,
       basicstyle=\ttfamily\footnotesize,
       numberstyle=\tiny,
       commentstyle=\color{gray},
       captionpos=b,
       frame=single,
       morekeywords={AGGREGATE,APPLY,CLEAR,FUNCTION,GROUPING,
         IS,MEASURE,OVER,TYPE,VISIBLE}}

% Uncomment the following line to show a list of all todos. (The todo
% and listoftodos commands are defined in the todonotes package.)
\listoftodos{}

\todo[inline]{Describe what are considered `visible' rows in a many-to-one
  join}

\todo[inline]{Before final submission, run this document through a
  spell-checker/grammar-checker; check that we are using backticks for
  open quotes; `window aggregates' not `windowed'; `evaluation
  context' rather than `filter context'.}

\section{Introduction}
\label{sec:introduction}

About thirty years ago the first Business Intelligence (BI) tools
were introduced. They had a semantic model based on the
multidimensional model, and good support for data exploration and
visualizations. Since then, the SQL language has expanded immeasurably
in its capabilities, adding support for XML, JSON, geospatial,
temporal, text and nested data.  An increasing proportion of business
data resides in powerful cloud SQL engines. But today’s BI tools
continue to use semantic models based on the multidimensional
model. Why?

Semantic models serve several purposes. They provide the building
blocks from which users can build queries (using some language,
perhaps graphical, perhaps textual). They guide users in the
construction of queries, and aid creation in visualizations and
reports. But we believe that their core strength is the ability to
express calculations in a concise manner, and to compose and reuse
those calculations.

In this paper, we show that the relational model imposes repetition of
filter expressions: changing the date range of a query requires updating
many \where{} clauses.
Therefore the challenge is how to extend the data model offered by SQL,
in ways that do not change the semantics of currently valid SQL expressions
or confound SQL users' expectations. Incorporating ideas from software
engineering, we extend SQL’s fundamental data type, the table, with a
new type of column, called a \emph{measure}, attaching a
\emph{context-sensitive expression} (CSE) to a table. We show that
tables with measures have a similar closure property to regular
tables.

SQL with measures can be expanded into traditional SQL. Therefore, the
path to integrating measures into existing systems is relatively
straightforward.

\subsection{Contributions}

\textbf{Encapsulation}. Measures define calculations close to the
data.  When a measure is used, it maintains its relationship to its
table.

\textbf{Clarity of query plan}. By eliminating the need for self-joins
and other forms of repetition in many queries, measures make it easier
for the optimizer to choose more efficient algorithms.

\textbf{Easier target for generative AI}. Generative AI algorithms
find it hard to correctly generate SQL queries that have repeated
subqueries, especially if those subqueries need to be
consistent. Measures enable more concise queries that are easier for
AI to generate.

\textbf{Modeling}. Measures allow you to define calculations in views,
and CSEs allow you to compose calculations into richer measures. SQL
can therefore take over work that was previously done in a BI tool
(semantic layer and multidimensional query language).

\textbf{Abstraction}. You can use a view without having knowledge of
the formulas in that view, or access to the tables referenced by the
view.

All of the above are delivered while retaining SQL’s closure
properties, security, and governance. Our extensions are backwards
compatible: queries that do not use measures have the same semantics
as regular SQL.

These extensions have been implemented in Apache Calcite
\cite{begoli2018calcite} and in Looker's Open SQL interface
\cite{looker2023openSql} (described further in subsection
\ref{ssec:looker}).

\section{Related work}

Adding measures to SQL requires us to bring together two theories, and
classes of database, long considered to be incompatible. It is worth
reviewing their parallel histories and the path to convergence.

The relational model \cite{codd1970relational} gave rise to relational
databases during the 1970s, and the dimensional model to
multidimensional databases in the 1990s. Vendors of the latter,
assisted by E.F. Codd \cite{codd1993beyond}, were at pains to point out
how relational databases' record-oriented storage was fundamentally
unsuitable for OLAP \cite{colliat1996olap}.

Multidimensional databases had no textual query language and were
generally inseparable from their user interface
(which was provided by the same vendor);
in early attempts to standardize access to multidimensional databases
(MDAPI \cite{olapCouncil1998mdapi} and JOLAP \cite{jsr2003jolap})
programmers would construct queries by calling an API.

It was difficult to imagine unifying relational and multidimensional
databases when they differed in the
fundamentals: whether there should be a query language; the data model
(relations, rows and columns versus cubes, dimensions, hierarchies,
and measures); and the algebra (relational operators select, project,
join, union versus dimensional operators slice, dice, drill, pivot
\cite{romero2007need, kuijpers2017algebra}).

Things began to change in the late 1990s. Kimball
\cite{kimballdata} introduced patterns to model complex business
analytics. In particular, semi-additive and non-additive measures
\cite{horner2004additivity} were patterns for measures more complex
than mere aggregate functions.  The SQL \cube{} operator
\cite{gray1997data} showed that it was possible to represent various
levels of subtotals in a query result without adding the complexity of
hierarchies to the data model. Analytic functions
(\overSql{}) \cite{zemke1999introduction} allowed running totals and
calculations of mixed grain, in some cases allowing the elimination of
self-joins \cite{zuzarte2003winmagic}.
\filter{}, \withinGroup{} and \withinDistinct{} clauses
\cite{hyde2021withinDistinct} provided finer control over the values
going into an aggregate function.

MDX was (at last!) a textual language for dimensional queries
\cite{whitehorn2004Mdx}. Unfortunately, its designers chose a syntax
that was superficially similar to SQL, and therefore many failed to
grasp its radically different semantics. Among those features were an
evaluation context consisting of one member for each of the current
cube's dimensions, and the ability to define calculated measures and
members using context-sensitive expressions. As a standard language,
there were multiple implementations of MDX, including Microsoft
Analysis Services, Mondrian \cite{back2013mondrian}, SAP BW, and
SAS. Some of these implementations were backed by relational databases
(a technique called ROLAP \cite{morfonios2007rolap}), and dimensional
languages came to be seen as a semantic layer on top of the relational
model.

The semantic layer's main contribution was not cubes.  (Data sets with
axes, hierachies and cells are harder for downstream tools to consume
than relations.)  It was the ability to define, just once, the
calculations central to the business, and to associate columns with
presentation metadata such as value formats and default sort
order. For example, Tableau's Level of Detail (LOD) expression
language allows users to control the grain at which aggregations
occur; Looker's centralized model makes governance easier and makes
calculations consistent.

\todo[inline]{But these operators must operate on the current data
  set.  They are confined by the relational model. If you wish to work
  on rows that have been removed by previous relational operators, you
  must use a join to add those rows back. Those joins require
  non-local edits to your query. The solution is MDX's
  context-sensitive expressions, but breaks the relational model.}

\todo[inline]{BI tools (Looker, Tableau LOD) start introducing
  dimensional concepts (grain, symmetric aggregates) on top of
  relational. Cubes and hierarchies were not necessary, and dimensions
  were just regular columns. The capabilities of the cloud SQL
  databases expanded, and responsibilities of the semantic layer began
  to shrink, to just something that expanded measures, and added
  visualization information (captions, format strings), perhaps
  governance and aggregate navigation.}

\todo[inline]{MDX results are dimensional (N axes, with cells at the
  intersection), which makes it difficult for relational tools to
  consume. Looker results are relational.}

\todo[inline]{MDX provides many things we don't need. Cell sets with N
  axes. Hierarchies. Measure as a dimension. Calculated members that
  are not measures. Lack of grain (you cannot ask a cube how many
  records it contains) and therefore inability to feed a cube or query
  result into relational operators.}

But these semantic layers' languages were not SQL; to benefit from a
semantic layer, users had to use its less-expressive, vendor-specific
query language. In the next section, we describe how to extend SQL so
that it can serve as the semantic layer.

\todo[inline]{But users were forced to choose between the semantic
  layer and SQL, a hard choice as data gravitated towards cloud SQL
  databases (Google BigQuery, Snowflake). Next, we describe how to
  extend SQL so that it can serve as the semantic layer.}

\section{Measures}

In this section we describe the new concepts and their SQL syntax. We
illustrate with examples; semantics are deferred to section
\ref{sec:semantics}.

\subsection{Tables are broken}

Tables are SQL's fundamental data model. Tables are implemented in
several ways, including base tables, views, and query specifications,
but any table you use in a query will have the same behavior. If you
have a query that uses a view, and you substitute a base table that
has the same rows as the view, the query will give the same
results. Furthermore, the model is closed: every SQL query returns a
table.

Tables are unable to provide reusable calculations. Suppose we have an
$\orders$ table that contains several orders for each product name and
customer (table \ref{table:orders}), and an expert SQL user has
written a query (listing \ref{lst:ordersByCatSql}) to compute the
profit margin for each product.

\begin{table}[ht!]
\centering
\begin{tabular}{|c|c|}
 \hline
 $\custName$ & $\custAge$ \\
 \hline\hline
 Alice  & 23 \\
 Bob    & 41 \\
 Celia  & 17 \\
 \hline
\end{tabular}
\caption{$\customers$ table}
\label{table:customers}
\end{table}

\begin{table}[ht!]
\centering
\begin{tabular}{|c|c|c|c|c|}
 \hline
 $\prodName$ & $\custName$ & $\orderDate$ & $\revenue$ & $\cost$ \\
 \hline\hline
 Happy & Alice & 2023/11/28 & 6 & 4 \\
 Acme  & Bob   & 2023/11/27 & 5 & 2 \\
 Happy & Alice & 2024/11/28 & 7 & 4 \\
 Whizz & Celia & 2023/11/25 & 3 & 1 \\
 Happy & Bob   & 2022/11/27 & 4 & 1 \\
 \hline
\end{tabular}
\caption{$\orders$ table}
\label{table:orders}
\end{table}

\begin{lstlisting}[caption=Summarizing $\orders$ by product name,
    label=lst:ordersByCatSql]
SELECT prodName,
  COUNT(*) AS c,
  (SUM(revenue) - SUM(cost)) / SUM(revenue)
    AS profitMargin
FROM Orders
GROUP BY prodName;
\end{lstlisting}

We now wish to create a SQL view that will allow less-expert users to
perform similar queries without typing out the formula for profit
margin. Listing \ref{lst:orderSummarySql} creates the view
$\summarizedOrders$ and attempts to use its $\profitMargin$ column in
a query to compute the profit margin for each product.

\begin{lstlisting}[caption=$\summarizedOrders$ view,
    label=lst:orderSummarySql]
CREATE VIEW SummarizedOrders AS
  SELECT prodName, orderDate,
    (SUM(revenue) - SUM(cost)) / SUM(revenue)
      AS profitMargin
  FROM Orders
  GROUP BY prodName, orderDate;

SELECT prodName, AVG(profitMargin)
FROM SummarizedOrders
GROUP BY prodName;
\end{lstlisting}

The query does not return the desired result; the desired result would
weigh each order equally, but actual result is an average over each
($\prodName$, $\orderDate$) combination. There is no correct query in
valid SQL; any correct query must read all rows in
$\orders$, but the rules of relational algebra do not allow the
$\summarizedOrders$ view to return any more information to a query
than its (summarized) rows.

\subsection{Measures and the \aggregate{} aggregate function}

To solve the problem, we introduce measures. Informally, a measure is
a column defined by a formula, and when that measure is used, the
formula is `copy-pasted' into the invocation. (More formally, as we
shall see later, a measure behaves as a context-sensitive
expression, taking its evaluation context from the clause in which it
is used.) This means each use of a measure can be expanded into an
traditional SQL query by explicitly, repetitively spelling out the
contextual filters.

Listing \ref{lst:enhancedOrdersSql} defines a view, $\enhancedOrders$,
that contains a measure, and uses it in a query.

\begin{lstlisting}[caption=$\enhancedOrders$ view,
    label=lst:enhancedOrdersSql]
CREATE VIEW EnhancedOrders AS
  SELECT orderDate, prodName,
    (SUM(revenue) - SUM(cost)) / SUM(revenue)
      AS MEASURE profitMargin
  FROM Orders;

SELECT prodName, AGGREGATE(profitMargin)
FROM EnhancedOrders
GROUP BY prodName;
\end{lstlisting}

There are a few things to note:
\begin{itemize}

\item The \asMeasure{} syntax indicates that $\profitMargin$ is to be
  a measure, not a regular column.

\item The $\enhancedOrders$ view does not contain a \groupBy{} clause,
  and therefore returns the same number of rows as the $\orders$
  table.

\item The measure formula contains aggregate functions, which would
  not be valid if this were a normal query. Measures need to be
  aggregatable --- that is, valid with any possible \groupBy{} clause
  in the query that uses it --- and therefore their formulas often
  contain aggregate functions.

\end{itemize}

The query uses the $\profitMargin$ measure and evaluates it in the
context of the current group row, aggregating over all rows with the
current value of $\prodName$. Users of the $\enhancedOrders$ view do
not need to know the formula for $\profitMargin$, nor need access to
the underlying $\orders$ table or its $\revenue$ and $\cost$ columns,
which meets our goal of providing an abstraction.

\subsection{Measures are not really aggregate functions}

The \aggregate{} function is present for largely cosmetic reasons.
SQL users know that a column that is not in the \groupBy{} clause must
be wrapped in an aggregate function when used in the \select{}
clause, so the \aggregate{} function makes such users (and tools that
generate SQL) more comfortable. As an aggregate function, \aggregate{}
conveniently converts any query into an aggregate query.

But framing measures as aggregate functions sells them short. They are
in fact evaluated very differently from aggregate functions. Consider
the following query (listing \ref{lst:queryEval}).

\begin{minipage}{\linewidth}
\begin{lstlisting}[caption=Evaluating a query,
    label=lst:queryEval]
SELECT prodName, AGGREGATE(profitMargin),
  COUNT(*)
FROM EnhancedOrders
GROUP BY prodName;

prodName profitMargin count
======== ============ =====
Acme     0.60         1
Happy    0.47         3
Whizz    0.67         1
\end{lstlisting}
\end{minipage}

What happens in the $\select$ clause as the
query is about to emit the row for `Happy'? \groupBy{} has assembled a
group of 3 rows for which $\prodName$ equals `Happy', and the
\countStar{} aggregate function is evaluated in the usual way over
these rows, emitting the value $3$.

The measure does not operate on the group rows (except indirectly).
Its only argument is the
evaluation context, which consists of the predicate\footnote{We have
simplified a little; if \prodName{} allowed null values, the
predicate would use \isNotDistinctFrom{}, rather than
\ssql{=}, in order to handle null values correctly.}
$$\sql{prodName\,=\,`Happy`}.$$

The effect is as if the query has been expanded as follows (listing
\ref{lst:expandedQuery}):

\begin{lstlisting}[caption=Query after expansion of measure,
    label=lst:expandedQuery]
SELECT prodName,
  (SELECT (SUM(i.revenue) - SUM(i.cost)) / SUM(i.revenue)
    FROM Orders AS i
    WHERE i.prodName = o.prodName),
  COUNT(*)
FROM Orders AS o
GROUP BY prodName;
\end{lstlisting}

The measure has been replaced by a scalar subquery. The subquery is
over \orders{}, the base table of the view in which the measure was
defined, and uses the same formula. To the subquery has been added a
\where{} clause that expresses the evaluation context, and therefore
the formula will be evaluated over the precise subset of rows in
\orders{}.

In the next section, we shall define measures in terms of
context-sensitive expressions.

\subsection{Context-Sensitive Expressions}

You might regard a measure as simply `a column that knows how to
aggregate itself,' and indeed many measures are just that. But the
goal is reusable calculations, which means that the client query does
not know the measure's formula, and the measure may use data that is
not accessible to the client.

So, we define the behavior of measures in terms of a new concept: the
\emph{context-sensitive expression}. Some definitions:

\begin{itemize}

\item A \textbf{context-sensitive expression} (CSE) is an expression
  whose value is determined by an evaluation context.

\item An \textbf{evaluation context} is a predicate whose terms are
  one or more columns from the same table.

\item This set of columns is the \textbf{dimensionality} of the CSE;
  we sometimes informally refer to these columns as \textbf{dimension
    columns} even though they are regular columns.

\item A \textbf{measure} is a special kind of column that becomes a
  CSE when used in a query. Its dimensionality is the set of
  non-measure columns in its table.

\item If a query references a table that has a measure, then any use
  of that measure in an expression has an \textbf{implicit evaluation
    context}. This context depends on the values of the measure's
  dimension columns and on the call site (which query clause, and
  whether there are joins or filters).

\item The \textbf{data type} of a CSE is \emph{t} \measure{}, for some
  data type \emph{t}; for example \ssql{INTEGER} \ssql{MEASURE}).

\item The \textbf{evaluation operator} \eval{} evaluates a CSE in the
  evaluation context of the call site; if the expression has type
  \emph{t} \measure{}, the value has type \emph{t}.

\item The \textbf{context transformation operator} \at{} modifies the
  evaluation context.

\end {itemize}

Applying these concepts to the query in listing
\ref{lst:queryEval}:

\begin {itemize}

\item The measure in the query is \profitMargin{}, and its
  dimensionality is the column set \{\prodName{},
  \orderDate{}\}.

\item \profitMargin{} has type \ssql{DOUBLE} \measure{}, and therefore
  \ssql{AGGREGATE(o.profitMargin)} has type \ssql{DOUBLE}.

\item \profitMargin{} is a measure, and therefore a reference to it is
  CSE.

\item \ssql{AGGREGATE(o.profitMargin)} expands to \\
  \ssql{EVAL(\-o.profitMargin} \at{} \ssql{(VISIBLE))}.\footnote{The
  \at{} operator and its \visible{} modifier will be explained in
  subsection \ref{ssec:atOperator}.}

\item The call site is the \select{} clause of an aggregate query, and
  therefore the evaluation context is a predicate that restricts to
  the rows matching the current group key, \prodName{} \ssql{=}
  \ssql{o.prodName}. Per the requirements of an evaluation
  context, it is in terms of one of \texttt{profitMargin}'s dimension
  columns, \prodName{}. (The right-hand side of the equality,
  \ssql{o.prodName}, is a correlation variable that is effectively
  constant when the predicate is invoked.)

\item Substituting the measure with a scalar subquery and a predicate
  that expresses the evaluation context yields the expanded query in
  listing \ref{lst:expandedQuery}, as expected.
\end {itemize}

CSEs and aggregate functions have fundamentally different evaluation
models:

\begin {itemize}

\item Aggregate functions, like relational algebra, are
  \emph{bottom-up}. The result of the calculation depends on the input
  rows, and the sequence of operators applied to them.

\item CSEs are \emph{top-down}. The result of the calculations is
  determined by the evaluation context.

\end {itemize}

\todo[inline]{The following section may be too verbose for this point
  in the paper. Maybe we can make the point concisely here, and
  revisit in discussion. The second point - the need for repeated
  subqueries and self-joins - could be better illustrated with an
  example in the section on AT.}

The top-down evaluation model has a number of advantages.

\begin {itemize}

\item Whereas aggregate functions can only be used in call sites where
  there is a set of rows to aggregate over, such as the \select{} or
  \having{} clause of a \groupBy{} query, measures and CSEs can be
  evaluated at any call site.

\item If you wish to evaluate a calculation in different contexts (say
  to compute profit growth between last year and this year, or to
  compare profit for a particular product with that for all products),
  top-down is more concise. In bottom-up, each calculation requires a
  separate pass over the input rows. In practice, this results in
  queries that have similar repeated subqueries and self-joins to
  combine the results of those subqueries on their common keys.

\item Top-down makes it easier to manage the grain of a calculation
  (daily versus monthly, per-order versus per-customer). A measure is
  locked to the grain of its defining table, and joining another table
  does not introduce double-counting the way it often does for
  bottom-up calculations.

\end {itemize}

\todo[inline]{Describe how measures are defined so as to be able to
  use evaluation context}

\todo[inline]{Describe what are the other kinds of context-sensitive
  expression, besides measures?}

\todo[inline]{Measures are a contract. AS MEASURE is not necessarily
  the only way to define them.}

\subsection{Modifying the evaluation context}
\label{ssec:atOperator}

In the previous section we saw that CSEs are evaluated in an
evaluation context that depends on the call site. We now introduce the
\at{} operator, which allows you to modify the evaluation
context. Syntax is as follows:

$$ cse\ \mathtt{AT}\ (modifiers)$$

where \emph{cse} is a context-sensitive expression and
\emph{modifiers} is a list of context modifiers as shown in table
\ref{table:atClause}. If there are multiple modifiers, they take
effect in sequence;
\emph{cse} \at{} (\emph{modifier\textsubscript{1}}
\emph{modifier\textsubscript{2}})
is equivalent to
(\emph{cse} \at{} (\emph{modifier\textsubscript{2}})) \at{}
(\emph{modifier\textsubscript{1}}).

\begin{table}[ht!]
\centering
\begin{tabular}{p{2.8cm}|p{5.2cm}}
 \hline
 Syntax & Effect \\
 \hline\hline

 \all{} & Sets the evaluation context to \ssql{TRUE} \\

 \all{} \emph{dimension} [\ \emph{dimension}... ] & Removes any
 \emph{dimension} terms from the evaluation context \\

 \set{} \emph{dimension} = \emph{expression} & Adds a \emph{dimension
 = expression} term to the context (replacing any occurrence of
 \ssql{CURRENT} \emph{dimension} with the current value of
 \emph{dimension}), removing any existing \emph{dimension} terms \\

 \visible{} & Adds terms to the evaluation context for the current
 query's \where{} clause and join conditions (if present), to ensure
 that measures are calculated over only the rows returned by the query
 \\

 \where{} \emph{predicate} & Sets the evaluation context to
 \emph{predicate} \\

 \hline
\end{tabular}
\caption{Context modifiers}
\label{table:atClause}
\end{table}

\todo[inline]{REVIEW: The `terms' and `dimension terms' concepts are a
  bit squishy. Should we define them more explicitly? If so, we should
  say that `CURRENT d' is only allowed if 'd' has already been set,
  and set to a singleton value. Does 'WHERE d1 = v1 AND d2 = v2' have
  the same effect as 'SET d1 = v1 SET d2 = v2'?}

\todo[inline]{REVIEW: Does WHERE replace all current filters
  (implicitly doing an ALL) or does it add?}

\todo[inline]{Give an example of measure-based filters (e.g. revenue
  from products that had more than 100 orders last year). And note
  that the measures are evaluated in the current context, therefore
  modifiers don't commute.}

\textbf{ALL}. The \all{} modifier allows you to compute a grand
total. For example, the following query (listing
\ref{lst:percentTotalQuery}) shows each product's revenue and its
proportion of the total revenue:

\begin{lstlisting}[caption=Query with proportion of total revenue,
    label=lst:percentTotalQuery]
SELECT prodName, sumRevenue,
  sumRevenue / sumRevenue AT (ALL prodName)
    AS proportionOfTotalRevenue
FROM (SELECT *,
    SUM(revenue) AS MEASURE sumRevenue
  FROM Orders) AS o
GROUP BY prodName;
\end{lstlisting}

When the query is emitting a row, the evaluation context for the
top-level \sumRevenue{} will be \prodName{} \ssql{=}
\ssql{o.prodName}, but due to the \at{} operator, the evaluation
context for the \sumRevenue{} measure inside the \ssql{sumRevenue\,
  AT\, (ALL\, prodName)} expression will be \ssql{TRUE}. The measure
\sumRevenue{} will be evaluated by iterating over the orders of a
particular product, whereas \ssql{sumRevenue\, AT\, (ALL\, prodName)}
will be evaluated by iterating over all orders.

\all{} with no arguments removes all filters, even filters not
associated with a particular dimension, and therefore the measure will
be evaluated over its entire source table.

\textbf{SET}. The \set{} modifier allows you to change the value of
one dimension. The following query (listing \ref{lst:queryEval2}) uses
\set{} with the \orderYearSql{} dimension to show profit margins in 2024 and
2023 for products sold in 2024:

\begin{lstlisting}[caption=Comparing profit margins in 2023 and 2024,
    label=lst:queryEval2]
SELECT prodName, orderYear,
  profitMargin,
  profitMargin AT (SET orderYear = CURRENT orderYear - 1)
    AS profitMarginLastYear
FROM (SELECT *,
    (SUM(revenue) - SUM(cost)) / SUM(revenue)
      AS MEASURE profitMargin,
    YEAR(orderDate) AS orderYear
  FROM Orders)
WHERE orderYear = 2024
GROUP BY prodName, orderYear;
\end{lstlisting}

This query is doing something novel for SQL: it is evaluating the
\profitMargin{} measure over data that has already been removed from
the query by the \where{} clause.

The \current{} qualifier applied to a dimension returns the null value
if the dimension has not been constrained to a single value by a
\set{} modifier or \where{} clause in the enclosing evaluation
context.

If the argument to \set{} (or \all{}) is an expression, such as
\ssql{DAYOFWEEK(orderDate)}, it is treated as an \emph{ad hoc}
dimension. \emph{Ad hoc} dimensions do not greatly complicate the
semantics for evaluating measures. All filters in the evaluation
context, whether on dimensions, or on expressions involving
dimensions, are combined into a single predicate, and the measure
value is only determined only by values returned by the predicate, not
the structure of the expressions that built that predicate.

\textbf{VISIBLE}. The \visible{} modifier adds terms to the evaluation
context so that the measure only includes rows allowed by the current
\where{} clause\footnote{And join conditions, as we shall see in
subsection \ref{ssec:measuresAndJoins}}. Consider the following query
(listing \ref{lst:visible}), which computes the count and sum of
revenue for orders not made by Bob, grouped by product.

% profitMargin for (sum(revenue) - sum(cost)) / sum(revenue)
\begin{lstlisting}[caption=Query with visible totals,
    label=lst:visible]
SELECT o.prodName,
  COUNT(*) AS c,
  AGGREGATE(o.sumRevenue) AS rAgg,
  o.sumRevenue AT (VISIBLE) AS rViz,
  o.sumRevenue AS r
FROM (SELECT *, SUM(revenue) AS MEASURE sumRevenue
  FROM Orders) AS o
WHERE o.custName <> 'Bob'
GROUP BY ROLLUP(o.prodName);

prodName   c  rAgg  rViz     r
======== === ===== ===== =====
Happy      2    13    13    17
Whizz      1     3     3     3
           3    16    16    25
\end{lstlisting}

Do we wish the grand total (the last row, with empty \prodName{}) to
include purchases by Bob, excluded by the \ssql{WHERE} clause?  There
are cases where each would make sense, and the \visible{} modifier
makes it possible to choose. The \ssql{r} column, which uses the
default evaluation context ignoring the \where{} clause, includes all
customers; \ssql{rViz}, which uses the \visible{} modifier, includes
only orders not made by Bob.

\countSql{} and \aggregate{} (columns \ssql{c} and \ssql{rAgg}) total
only the visible rows, as is customary for SQL aggregate
functions. This is why we remarked earlier that
\texttt{AGGREGATE(}\emph{m}\texttt{)} expands to
\texttt{EVAL(}\emph{m}\ \texttt{AT\,(VISIBLE))} for any measure
\emph{m}.

\todo[inline]{Explain that the WHERE clause in a subquery that defines
  a measure is baked into the measure and cannot be subverted.}

\textbf{Advanced context modifiers}. We do not regard the list of
modifiers allowed by the \at{} operator as complete or final. For
instance, there is a compelling argument for `named filters' that can
be added by a UI control and removed or overridden in the evaluation
context by the SQL runtime, but we have not included them in this
paper. The reason is simple: when a measure is evaluated, it cares
only about the predicate --- do I include this row in the total, or
not? --- and not about the structure of the evaluation context that
created the predicate.

We look forward to useful context modifiers devised by others, and we
believe that they will not change the fundamentals of how measures are
evaluated.

\subsection{Measures and joins}
\label{ssec:measuresAndJoins}

It's worth discussing how measures work in join queries, because
people's desired semantics are complicated, and because the natural
semantics of measures is different --- we believe in a good way ---
from people's expectations of SQL, namely aggregate functions.

Consider a query that joins a table with measures (\enhancedCustomers{})
to another table (\ssql{Orders}).

\begin{lstlisting}[caption=Joining measures,
    label=lst:measureJoin]
WITH EnhancedCustomers AS
  (SELECT *,
      AVG(custAge) AS MEASURE avgAge
    FROM Customers)
SELECT o.prodName,
  COUNT(*) AS orderCount,
  AVG(c.custAge) AS weightedAvgAge,
  c.avgAge AS avgAge,
  c.avgAge AT (VISIBLE) AS visibleAvgAge
FROM Orders AS o
  JOIN EnhancedCustomers AS c USING (custName)
WHERE c.custAge >= 18
GROUP BY o.prodName;
\end{lstlisting}

The join is one-to-many. A customer may match zero, one or many orders.
The query semantics do not depend on the SQL system knowing
which primary keys and foreign keys exist. That would arguably
contradict the data independence principles of SQL.

How many rows are returned? What are the values of \ssql{prodName} and
\ssql{orderCount}? These are straightforward questions to answer,
because measures do not affect the basic operations of SQL, such as
the number of rows in a relation. A row is returned for each product
that has at least one order to a customer 18 or older, and the
count is the number of orders.

The \ssql{weightedAvgAge} column computes the average customer age in the
traditional SQL way. It joins orders to customers, removes
customers under 18, and for all joined rows with the same product computes
a weighted average of the ages. If
one product has one order, and another has two orders from the same customer,
the second contributes twice as much to the average as the first.

Which average is correct --- the weighted average, the visible average
(containing customers only 18 or older), or the unweighted average ---
depends, of course, on what you want the number for, but it is useful
that there is a concise syntax for each.

\section{Semantics}
\label{sec:semantics}

\todo[inline]{Back away from `function values' language. Function
  values are worrying to database theorists because they can cause
  queries to not terminate.}

In the previous section, we introduced several new SQL concepts:
measures, context-sensitive expressions, and operations that modify
the evaluation context. We now define their semantics.

In our data model for analytic SQL, which adds measures to tables, it
is important to separate how measures are defined from how they are
used. A measure \emph{may} be defined using the \asMeasure{}
construct, or it may be defined in some other way, but any query that
uses that measure should never be able to tell.

To keep the semantics separate, we therefore proceed as follows. First
we define the evaluation context, and how it is perceived by
expressions. Then we define how a table interacts with the query
optimizer to convert measure references into expressions. Lastly, we
define the \at{} operator.

\subsection{Lambdas}

In order to simplify the explanation of semantics, we use a
functional extension to the SQL language, as follows.

\textbf{A note on safety}. Adding function values, also
known as closures or lambdas, to SQL would make
the language Turing complete, and therefore make it difficult to
reason about query termination. This proposal does not step into those
stormy waters. First, these extensions are expanded for the
query optimizer. We do not make them accessible to the SQL user. The
use of closures here is just for clarity of exposition, particularly
to clarify which definition is meant when a name is defined in
different scopes. Second, the closures that we
introduce during the planning process are gone before planning is
complete. There are no function values at runtime.

\begin{itemize}

\item A \textbf{closure} represents a function expression.
  Its type is
  $$\texttt{FUNCTION($A$) RETURNS $R$},$$ where \ssql{FUNCTION} is a type
  constructor, $A$ is the argument type and $R$ is the result type.

\item A \textbf{lambda} (denoted \ssql{->}) is a SQL operator that
  denotes a closure. For example, $$\sql{(x: INTEGER) \rightarrow
    MOD(x, 2) = 0}$$ is a function expression that returns whether its
  integer argument is even; its type is \texttt{FUNCTION(INTEGER)
    RETURNS BOOLEAN}.

\item \textbf{\texttt{APPLY}} is a SQL operator that applies a closure
  to an argument. For example, $$\sql{APPLY((x: INTEGER) \rightarrow
    MOD(x, 2) = 0, 3)}$$ returns \ssql{FALSE}, because 3 is not even.

\end{itemize}

\subsection{Semantics of context-sensitive expressions}
\label{ssec:rewriteAlgorithm}

Having defined lambdas, we outline a process to rewrite measures.

\begin{itemize}

  % Why are auxiliary functions needed? Why not have the measure
  % column simply return a closure for each row, that could be invoked
  % directly? That would give no way of invoking the measure on an
  % empty group --- if the table is empty, or if a measure is invoked
  % in the manner of a window aggregate function over an empty window
  % --- and it would give no guarantee that the formula for the
  % measure is consistent across all rows in a group.

\item For every measure $M$ of value type $V$ that belongs to a table
  whose row type (excluding measures) is $R$, the system defines an
  \textbf{auxiliary function} that has name $\mathtt{compute}M$%
  \footnote{Or a variation of that name that is unique within the
  namespace}
  and type \texttt{FUNCTION(rowPredicate: FUNCTION($R$) RETURNS
    BOOLEAN) RETURNS $V$}. The auxiliary function must be pure and
  deterministic but may contain a reference to the table.

\item At any point in the query where $M$ is accessible, the system is
  able to generate a \textbf{row predicate} of type
  \texttt{FUNCTION($R$) RETURNS BOOLEAN}. The row predicate reflects
  the evaluation context of the measure.

\item If an expression occurs within a call to \at{}, the evaluation
  context is modified by applying the modifiers in succession.

\item From a evaluation context for $M$ can be generated a \textbf{row
  predicate} of type \texttt{FUNCTION($R$) RETURNS BOOLEAN}

\item At any point in the query where $M$ is referenced in an
  expression, the compiler replaces the measure reference with a call
  to its auxiliary function; the argument is the row predicate and the
  return value has type $V$, as required.

\end{itemize}

\todo[inline]{Suppose a query is a join, and has access to several
  measures from each side of the join. The evaluation context will
  have the dimensionality of the measure in question. So it will be
  different for measures from each side of the join. We should explain
  this, and give an example.}

Here is an example that follows the above rules. We have the following
query (listing \ref{lst:semanticsExample}) that computes the ratio of
this year's revenue to last year's revenue, for each product.

\begin{lstlisting}[label=lst:semanticsExample,
    caption=Year over year revenue by product]
CREATE VIEW OrdersWithRevenue AS
  SELECT *, SUM(revenue) AS MEASURE sumRevenue
  FROM Orders;

SELECT prodName, YEAR(orderDate) AS orderYear,
  sumRevenue / sumRevenue AT
     (SET orderYear = CURRENT orderYear - 1) AS ratio
FROM OrdersWithRevenue
GROUP BY prodName, YEAR(orderDate);
\end{lstlisting}

The measure $M$ is \sumRevenue{}, and the row type $R$ is the type
\ssql{OrdersRow} consisting of the non-measure columns of the
\orders{} view. Listing \ref{lst:semanticsExpanded} shows
the definition of a type for $R$, and the query with the two
references to \sumRevenue{} replaced by calls to the auxiliary function
\computeSumRevenue{}. Each call has a row predicate that reflects the
evaluation context at its call site. The first call has the evaluation
context of output from the \groupBy{}; in the second call, the year
in the filter context is set to the year before the current one.

\begin{minipage}{\linewidth}
\begin{lstlisting}[label=lst:semanticsExpanded,
    caption=Expansion of query comparing average revenue]
-- Row definition
CREATE TYPE OrdersRow AS ROW (prodName: VARCHAR,
    custName: VARCHAR, orderDate: DATE,
    revenue: INTEGER, cost: INTEGER);

-- Auxiliary computation for sumRevenue
CREATE FUNCTION computeSumRevenue(
    rowPredicate: FUNCTION(r: OrdersRow)
      RETURNS BOOLEAN) AS
  SELECT SUM(o.revenue)
  FROM Orders AS o
  WHERE APPLY(rowPredicate, o);

-- After expansion of sumRevenue occurrences
SELECT o.prodName, YEAR(o.orderDate) AS orderYear,
   computeSumRevenue(
      r -> r.prodName = o.prodName AND
           YEAR(r.orderDate) = YEAR(o.orderDate))
  / computeSumRevenue(
      r -> r.prodName = o.prodName AND
           YEAR(r.orderDate) = YEAR(o.orderDate) - 1)
  AS ratio
FROM Orders AS o
GROUP BY prodName, YEAR(orderDate);
\end{lstlisting}
\end{minipage}

\todo[inline]{Specify the row predicate for every clause and other
  location in the query.}

%\section{Implementation}
%\subsection{External measures}
%\subsection{Implementing SQL measures}
%\subsection{Flattening the scalar subqueries that result from measure expansion}

\section{Discussion}

\subsection{Self-joins and window aggregates}
\label{ssec:selfJoins}

There is a fascinating correspondence between measure expressions,
window aggregates, and self-joins.

The correspondence between window aggregates and self-joins (expressed
in the from of correlated subqueries) was first noted in
\cite{zuzarte2003winmagic}, whose WinMagic algorithm rewrites certain
kinds of subquery to window aggregates. The four queries in listing
\ref{lst:winMagic} are equivalent, and all find orders whose revenue
is higher than the average for their product. WinMagic provides an
algorithm to rewrite query 1 (correlated subquery) to query 3 (window
aggregates); queries 2 and 4 are equivalent queries using self-join
and measures.

\begin{minipage}{\linewidth}
\begin{lstlisting}[label=lst:winMagic,
    caption=Four equivalent queries to find orders with more revenue
        than average for their product]
-- Query 1: correlated subquery
SELECT o.prodName, o.orderDate
FROM Orders AS o
WHERE o.revenue >
  (SELECT AVG(revenue)
    FROM Orders AS o1
    WHERE o1.prodName = o.prodName);

-- Query 2: self-join
SELECT o.prodName, o.orderDate
FROM Orders AS o
LEFT JOIN
  (SELECT prodName, AVG(revenue) AS avgRevenue
    FROM Orders
    GROUP BY prodName) AS o2
    ON o.prodName = o2.prodName
WHERE o.revenue > o2.avgRevenue;

-- Query 3: window aggregate
SELECT o.prodName, o.orderDate
FROM
  (SELECT prodName, revenue, orderDate,
      AVG(revenue) OVER (PARTITION BY prodName)
        AS avgRevenue
    FROM Orders) AS o
WHERE o.revenue > o.avgRevenue;

-- Query 4: measures
SELECT o.prodName, o.orderDate
FROM
  (SELECT prodName, orderDate, revenue,
      AVG(revenue) AS MEASURE avgRevenue
    FROM Orders) AS o
WHERE o.revenue >
    o.avgRevenue AT (WHERE prodName = o.prodName);
\end{lstlisting}
\end{minipage}

Observe that queries 3 and 4 have very similar structure. This is
because the \overSql{} operator (window aggregation) and \at{}
operator (measures) have the same function: to evaluating a
calculation over a collection of rows meeting some criterion. \at{} is
more powerful than \overSql{}; it can evaluate arbitrary predicates
where \overSql{}'s \partitionBy{} can evaluate only \ssql{=}
predicates; and it can query rows that have been removed by a \where{}
clause.

Why is the WinMagic rewrite beneficial? Observe that \orders{}
appears twice in queries 1 and 2 but only once in 3 and 4. This
suggests to the optimizer an execution strategy that you might call
`localized self-join'. The engine scans order records grouped by product;
when it has finished a product, and knows the average revenue
of that product, it rewinds to the beginning of the product and
emits orders whose revenue is greater than the average.

This strategy, of small loops probing into intermediate results cached
in memory, is characteristic of in-memory OLAP engines. We believe it
is worth investigating whether this strategy is also beneficial in SQL
engines.

Aside from the runtime benefits, the queries with less repetition are
easier to optimize, because optimizers have difficulty identifying
common sub-trees in relational algebra.

\subsection{Hierarchies}

We chose not to explicitly support hierarchies.
Hierarchies are a major part of dimensional
systems, but they complicate the language and are largely used for
user interface concerns (for example, suggesting fields to drill down
on). For our purposes, it is sufficient to be able to treat any
expression on a dimension (for example, \ssql{YEAR(orderDate)}) as an
\emph{ad hoc} dimension.

That said, when I set the \texttt{year} dimension, I should not have
to explicitly clear the \texttt{month} dimension. In order to achieve
that effect, we hope (in a future version of this language) to allow
dimensions to be `linked' for purposes of their \all{} and \set{}
behavior.

\todo[inline]{% Do we want this?
  \textbf{Relationships}. SQL offers a limited recognition of
  relationships with its \ssql{natural\,join}
  syntax. Context-sensitive expressions should never aggregate over
  the Cartesian product of columns coming from different tables, which
  refer to the same semantic concept.}

\subsection{Wide tables}
\label{ssec:wideTables}

Business Intelligence tools typically have a `cube' or `business view'
concept that contains measures from a fact table and columns from
several dimension tables. This is attractive to end-users because they
do not need to specify joins. Without measures, `wide tables' composed
as join views were not advisable because denormalization would
introduce inconsistencies such as double-counting. But
with measures, calculations maintain their own consistency, and wide
tables are a recommended practice.

Wide tables can also contain measures with complex behaviors:

\begin{itemize}

\item A \textbf{semi-additive measure} rolls up using different
  aggregate functions on different dimensions but can sometimes be
  summed; for example, an \emph{items on hand} (inventory) measure
  rolls up using \lastValue{} on the time dimension and
  \ssql{SUM} on other dimensions;

\item A \textbf{non-additive measure} never aggregates by summing,
  typically a calculation based on other measures; for example,
  \emph{return rate} is the ratio of product units sold to product
  units returned.

\item Other custom measures might use a different formula for
  different levels of a hierarchy; for example, the \emph{revenue}
  measure might have a different formula at a business unit level than
  at a country level. The SQL \groupingId{} function can be used
  to identify the level.

\end {itemize}

\subsection{Composability}

Measures are composable in several ways.

First, as we have mentioned, the query language is closed. A query can
reference tables with (or without) measures, and returns a table with
(or without) measures. Queries can therefore be nested to arbitrary
depth, as in regular SQL. Views with measures can be created upon
relations (such as a traditional relational database, or a directory
of CSV files) that do not have measures.

Second, measures can reference measures in the same query. Measures
defined using the \asMeasure{} syntax can reference by name other
measures defined in the same \select{}.

(We do not, in the current language, allow recursive or mutually
recursive measures. We believe that they are useful, but there are
implementation hurdles. Termination is one concern, although
spreadsheet formulas manage perfectly well without provable
termination. A greater concern is that recursive measures cannot be
implemented using a static rewrite, and will require some form of
unbounded state, such as a call stack.)

Third, a measure can reference a measure or measures from an input
table, and thus a measure seems to be propagated effortlessly through
a stack of nested queries. But the semantics are defined one step at a
time. Each query is evaluating a context-sensitive expression, in its
own evaluation context, and defining a new measure whose
dimensionality is determined by the columns that it projects, and
that new measure is consumed by its enclosing query.

\subsection{Security}

SQL's security model is simple and robust: if I own
tables that contain sensitive information, I can write a query that
accesses those tables, publish that query as a view, and grant access
to the view but not the underlying tables. People can access the data
that I allow them to, and the optimizer will ensure that those queries
have efficient plans.

Do views with measures offer the same robust security model? The
answer is yes. This may be surprising, given that measures return much
richer values than regular columns, so let's justify that assertion.

A regular SQL view, without measures, returns a fixed amount of
information; this is easy to see because if I replace the view with a
base table with the same contents, every possible query will return
the same results.

Now consider a view that has regular columns \emph{a} and \emph{b},
hidden columns \emph{c} and \emph{d} that are not projected by the
view, and measures \emph{m} and \emph{n}. Queries that only use
\emph{a} and \emph{b} are straightforward; they map to the relational
core. But what of queries that also use the measures? They too are
bounded. Each measure does not return a single value, of course, but
it returns a map that can be read by providing a predicate. If I ask
for the value of measure \emph{m} with the predicate \emph{a = 0 and b
< 10}, it returns 6; if I ask for the value of \emph{m} with the
predicate \emph{a = 1}, it returns 12, and so forth.

Furthermore, the predicate can only be in terms of the dimension
columns \emph{a} and \emph{b}, not in terms of the hidden columns
\emph{c} and \emph{d}. If two rows in the underlying table(s) cannot
be distinguished based on their \emph{a} and \emph{b} values, then I
cannot construct a predicate to separate them.

To use an analogy, if regular column values are like pixels of a
discrete image, then measures are like holograms; their data has more
dimensions, but is still finite.

A view with measures thus allows me to create an interface that limits
which questions can be asked of the underlying data.

\subsection{Looker's Open SQL Interface}
\label{ssec:looker}

Looker\cite{looker} is a BI platform that was acquired by Google in
2019 and is now part of Google Cloud. Using Looker's
LookML\texttrademark{} language, analysts define objects called
``Explores'', which are a form of the wide tables described in
subsection \ref{ssec:wideTables}. These are the starting point for
data exploration via pivot tables, charts, and dashboards.

Looker also serves as a semantic layer for third-party visualization
tools such as Google Sheets, Microsoft Power BI, Tableau,
and ThoughtSpot. Those
tools query the Explores, benefiting from the joins, measures and
other calculations, and presentation and navigation information
encapsulated in them. Organizations choose to use a semantic layer so
that Explores are defined just once, in one place, as opposed to many
redundant and inconsistent definitions in the visualization layer.

In Looker's Open SQL Interface\cite{looker2023openSql}, each Looker
Explore appears as a SQL table, the measures in that Explore appear as
measure columns, and the dimensions in that Explore appear as regular
columns. The SQL Interface accepts SQL queries that adhere to
GoogleSQL syntax, and supports most of the BigQuery operators.

Before the SQL Interface was introduced, building a connector from a
third-party tool was complicated, because expressions in the tool's
expression language had to be translated into Looker's expression
language. Connectors built using the SQL Interface are much simpler,
and are similar to the tools' existing connectors to conventional SQL
databases. When generating SQL, tools can use measures defined in
Looker (e.g. \ssql{AGGREGATE(profitMargin)}) or can define their own
measures using aggregate functions on top of regular columns
(e.g. \ssql{SUM(revenue)}).

The implementation uses Apache Calcite's SQL parser, query planner,
and SQL function library.

\subsection{Natural Language to SQL}
\label{ssec:generativeAi}

For applications such as natural-language-to-query translation,
including those powered by Large Language Models (LLMs) and Generative
AI, SQL-with-measures is an attractive target language, for three
reasons.

First, it manages complexity. Like humans, generative AI has
difficulty correctly generating large expressions, especially when
consistency is required between regions of those expressions that are
widely separated. If the target language is regular SQL, the generated
queries are large, deeply nested, and have many joins, including
complex self-joins. In SQL-with-measures, the joins and calculations
can be encapsulated in a view, and context-sensitive expressions
eliminate the need for self-joins, and therefore the generated query
is more concise and less complex.

Second, current query-generation systems use a multidimensional
semantic layer --- for example, Analyza \cite{dhamdhere2017analyza}
uses a catalog containing ``additional information about the type of
the column (e.g. is it a metric, dimension, etc.), data formats
(e.g. should the number be formatted as a dollar amount), and date
range defaults'' --- and measures allow us to encapsulate that
semantic layer as SQL views.

Last, the corpus of queries in SQL is larger than in any other query
language, and therefore training LLMs is much easier.

Early indications are that the generated queries are smaller and more
accurate --- and easier to understand.  More research is needed in
this area.

\section{Future work}

\subsection{Formal semantics}

In this paper, we have presented an informal semantics. It would be
useful if a future publication described a formal semantics for
measures, context-sensitive expressions and the evaluation
context. Perhaps these could be extensions to relational algebra.

The semantics of the \at{} operator should be clarified. It seems
reasonable to allow expressions within \set{}, for example
\ssql{profitMargin AT (SET YEAR(orderDate) = CURRENT YEAR(orderDate) -
  1)}.

The \current{} operator should return a valid value if the evaluation
context implies a single value for all possible rows; for example, if
the query has \ssql{GROUP BY FLOOR(orderDate TO MONTH)} all rows in a
given group will have the same month and therefore the same year. To
allow the SQL semantic analyzer to safely make that deduction, we need
new rules for deducing functional dependencies among expressions,
perhaps a notion similar to Calcite's nested time frames
\cite{hyde2022timeFrame}.

\todo[inline]{In discussion, describe problems of nondeterministic query. If
  expanding a query causes relations to be used more than once, and if
  those relations' results are nondeterministic, then the two
  evaluations may be inconsistent.}

\todo[inline]{In discussion, describe the problems accessing a measure in a
  context that requires access to a particular row. Namely the where
  clause, and window aggregates. If the table that defines the measure
  has a primary key, we're OK. (A ROWID function helps, but only if it
  maps to a primary key; if there is no such key, then generating a
  key will bump into the problems of nondeterminism.)}

\subsection{Generating queries from natural language}

As mentioned in subsection \ref{ssec:generativeAi}, research should
ascertain whether SQL-with-measures is an effective target language
for AI-powered query generation.

\subsection{Operators for managing grain}

Measure have the useful property that they preserve grain in the
presence of joins (preventing double-counting), but we need
more operators for managing grain.

For example, an \emph{items on hand} semi-additive measure might take
the count of each product on the warehouse shelf on the last day of
the time period, and then sum over all products and warehouses. A
\emph{rank change} non-additive measure might rank each product by
revenue in a given region and time period, and then compute the
difference with the rank in the previous time period. Such measures
perform multiple aggregation steps, each step using a different
aggregate function and occurring in a particular order.

A promising candidate is the \ssql{PER} clause for aggregate
functions, proposed as a generalization of Calcite's \withinDistinct{}
clause \cite{hyde2021withinDistinct}.

\subsection{Implementation strategies}

Strategies to implement queries with measures and context-sensitive
expressions require further study.

One strategy is to rewrite queries in terms of simpler operations. Our
algorithm in subsection \ref{ssec:rewriteAlgorithm}, which rewrites a
measure reference as a correlated scalar subquery, is general-purpose
but not very efficient. In simple cases (such as a query with
\groupBy{} and no \join{}) it may be valid to inline the measure
definition. In cases with joins, a \withinDistinct{} clause may be
introduced to preserve the measure's grain. The correspondences noted
in subsection \ref{ssec:selfJoins} suggests that some queries can be
rewritten to window aggregates, especially if window aggregates are
generalized to computations to access ``lost'' rows.

As we remarked earlier, recursive measures cannot be solved using a
static rewrite, and may require a new physical algorithm. That
algorithm may also be applicable to other cases.

\subsection{Forecasts and time series}

Forecasting and time series analysis are similar domains. Time series
analysis often involves interpolation, such as changing the temporal
grain of a measure (resampling) to match other measures, or to fill
gaps where no measurement is available; forecasting generally
extrapolates, creating estimates of a measure in the future based on
past values of that measure and related measures.

Both make extensive use of statistical techniques; for example,
autoregressive integrated moving average (ARIMA) can detect and
compensate for periodicity. Measures can simplify things for users: an
expert defines the calculations, encapsulates them in a model (view)
as measures, and the user can use the model without worrying about the
complexity.

A challenge to be solved is that both techniques create new values for
dimensions (for example, recording a revenue of zero on a holiday,
when the business is closed, or generating a revenue forecast for a
future year, for which there are not yet any orders). We will need to
devise a query syntax for synthesizing rows. At the same time, we can
answer the important question, ``How can I evaluate a measure on a
table that has no rows?''

\subsection{Log files and sequential processing}

Much modern data processing, especially during load and transformation
phases, takes place on log files that have a nested structure. Records
are processed in sequence, often in a single pass, but with a
processing context that includes the current record, sibling records
that occur within the same parent (such as the group of records for
the same browser session), the parent record, and perhaps other data
values computed from various ``ancestor'' records. Measures might
allow such calculations to be expressed declaratively.

On the related topic of sequence data, measures may be helpful in
organizing the complex rules for identifying logical business events
as part of the data model. Their relationship with SQL's existing
\matchRecognize{} clause \cite{zemke2007pattern} should be
investigated.

\section{Summary}

Measures are a natural extension to the relational data model. They
allow calculations, including aggregate functions, to be encapsulated
in the definition of a table. These calculations offer
context-dependent views of the table; not a single static image but
one that varies based on the viewer, like a hologram.

The evaluation context of a measure is established in its definition
and can optionally be adjusted when it is used, by making changes to
just the expression that invokes the measure.  This locality of
reference allows queries to be written concisely, allows queries to be
composed reliably, and brings modularity to relational systems using
SQL.

Recent explorations with LLMs remind us how challenging were those
non-local transformations that we previously required of human SQL
authors. Measures make these repetitive filters and self-joins
invisible, and we hope that they improve the lot of humans and
machines alike.

\section{Acknowledgments}

This work would not have been possible without many design discussions
and much patient, constructive feedback.  The authors would like to
thank their Google colleagues
Adam Wilson,
Alexey Leonov-Vendrovskiy,
Bengu Li,
David Wilhite,
Goetz Graefe,
Jeff Shute,
Lloyd Tabb,
Marieke Gueye,
Matthew Brown,
Mosha Pasumansky,
Riccardo Muti,
Romit Kudtarkar, and
Serhiy Tykhanskyy.

\bibliographystyle{ACM-Reference-Format}
\balance{}
\bibliography{bibliography}

\appendix

\end{document}